# Dynamically encircling an exceptional point in anti-*PT*-symmetric systems: asymmetric mode switching for symmetry-broken states

Xu-Lin Zhang,[1,2] Tianshu Jiang,[1] Hong-Bo Sun,[2,3] and C. T. Chan[1]

[1]*Department of Physics, The Hong Kong University of Science and Technology, Clear Water Bay, Hong Kong, China*
[2]*State Key Laboratory of Integrated Optoelectronics, College of Electronic Science and Engineering, Jilin University, Changchun, China*
[3] *State Key Laboratory of Precision Measurement Technology and Instruments, Department of Precision Instrument, Tsinghua University, Haidian, Beijing 100084, China*
(Dated: June 20, 2018)

Dynamically encircling an exceptional point (EP) in parity-time (*PT*) symmetric systems shows an interesting chiral dynamics, leading to asymmetric mode switching in which the output modes are different when the encircling direction is reversed. Here we show that the dynamical encircling of an EP in anti-*PT*-symmetric systems can also result in chiral dynamics if the starting/end point lies in the *PT*-broken phase, in contrast to *PT*-symmetric systems where chiral dynamics emerges if the starting/end point lies in the *PT*-unbroken phase. For many applications, such as signal processing using waveguides, the asymmetric mode switching of symmetry-broken modes in anti-*PT*-symmetric systems is more useful since each eigenmode is localized in one waveguide only. We develop an analytic theory for anti-*PT*-symmetric chiral dynamics and perform experiments using three waveguides to demonstrate the asymmetric mode switching. The new wave-manipulation phenomena observable in anti-*PT*-symmetric systems may pave the way towards designing on-chip optical systems with novel functionalities.

Non-Hermitian systems obeying parity-time (*PT*) symmetry, i.e., $[PT,H]=0$ with $H$ being the non-Hermitian Hamiltonian, have attracted considerable attention in recent years [1-4]. Most of the interesting properties of non-Hermitian systems are found at the degeneracy, known as exceptional point (EP) [5,6], which has led to many novel phenomena and plausible applications [7-17]. The topological structure of energy Riemann surface around the EP is of great interest and it is known that adiabatically encircling an EP can result in an intriguing "flipping of the eigenstate" phenomenon in which an eigenstate does not come back to itself after going around a loop in parameter space [18]. This phenomenon was demonstrated experimentally [19-21] by measuring the spectra and eigenfields at different points of an encircling loop. In contrast to these 'static' measurements in which the results at different locations are independently measured, dynamically encircling an EP where the phase information at different points is closely related is predicted to exhibit an intriguing chiral behavior [22-27] because of the non-Hermiticity induced non-adiabatic transitions (NATs) [28]. The chiral behavior was recently observed experimentally in microwave waveguides with suitably designed boundaries and losses [29]. In the experiment, the starting/end point of the loop lies in the *PT*-symmetric phase where the eigenmodes are symmetric and anti-symmetric modes (i.e., symmetry-unbroken modes). The observed chiral dynamics can be applied to asymmetric mode switching such that the output modes are different when the propagation direction in the waveguide is reversed. This unusual way of mode switching is unique to the *dynamical* encircling of the EP and has great potential for on-chip applications such as the design of nonreciprocal devices [30]. From the viewpoint of applications of on-chip devices using waveguide configurations, symmetry-broken modes (i.e., each eigenmode is localized in one waveguide only) are of more interest since they are typically used as the input and output of the system [31,32]. However, our recent work demonstrated that when the starting point of the loop lies in the *PT*-broken phase where the eigenmodes are symmetry-broken modes, dynamically encircling the EP results in a non-chiral transmission behavior [33] in the sense that the output modes are the same for both propagation directions. Therefore, the symmetry-broken modes cannot be used for asymmetric mode switching in *PT*-symmetric systems.

Anti-*PT*-symmetric systems, whose Hamiltonians obey $\{PT,H\}=0$, have also attracted a lot of attention recently [34-38]. From a mathematical point of view, the anti-*PT*-symmetric Hamiltonian can be obtained by multiplying the *PT*-symmetric Hamiltonian by a constant '*i*'. However, from a physical point of view, it is challenging to construct a realistic system that possesses anti-*PT* symmetry because it requires the coupling between the two bare states to be a purely imaginary value. As we will see, one way to satisfy this difficult requirement is to couple the two bare states through a third state [36,37]. Anti-*PT*-symmetric systems are shown to possess EPs as well. What would the dynamics be if the EP in anti-*PT*-symmetric systems is dynamically encircled? Is there new physics that can be employed for new applications unique to anti-*PT*-symmetric systems? These questions remain open.

In this work, we report the first experiment on the dynamical encircling of an EP in anti-*PT*-symmetric systems. We study an anti-*PT*-symmetric system consisting of three waveguides with an absorber attached to the middle one. The two gap distances are designed to vary continuously along the waveguiding direction so that the transmission of electromagnetic waves through the system is equivalent to a loop enclosing an EP in the parameter space. We discover a chiral transmission behavior so that different encircling directions result in different output states, when the starting/end point of the loop is designed to be in the *PT*-broken phase. As a result, the symmetry-broken states can be used for asymmetric mode switching in anti-*PT*-symmetric systems, whereas such application cannot be achieved in *PT*-symmetric systems. We propose a theoretical model to prove the chiral dynamics. We also perform experiments to demonstrate the asymmetric mode switching by measuring the transmission spectra and electric field distributions at microwave frequencies.

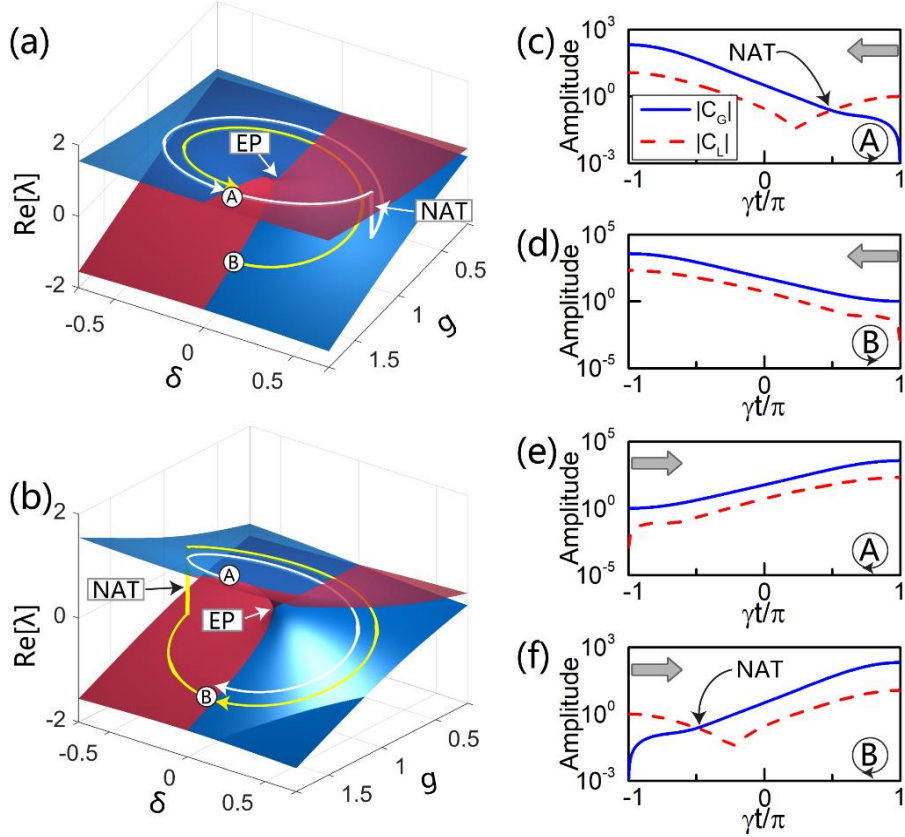

FIG. 1. (a) Real part of the eigenvalues as a function of $g$ and $\delta$. The white and yellow curve represent the trajectory for counter-clockwise loops with A and B as the initial state, respectively. (b) Same as (a) except for clockwise loops. (c)-(f) Calculated amplitudes of the eigenstates in the encircling process with different encircling directions and initial states (indicated in the inset). The grey arrows mark the starting points of the loops.

We start by investigating the dynamical encircling of an EP in a two-state system governed by $i\partial_t |\psi(t)\rangle = H(t)|\psi(t)\rangle$, where $|\psi(t)\rangle = [a(t), b(t)]^T$ is the state vector at a certain time $t$. The model Hamiltonian takes the form

$$H(t) = \begin{pmatrix} -g(t)+i\delta(t) & i\kappa \\ i\kappa & g(t)-i\delta(t) \end{pmatrix}, \quad (1)$$

where $g$ and $i\kappa$ denote the amount of detuning and coupling, respectively. The system is anti-*PT*-symmetric when $\delta = 0$. Without loss of generality, we set $\kappa = -1$ in the following analysis. We first calculate the eigenvalues $\lambda$ of the non-Hermitian system as a function of $g$ and $\delta$ and show the real parts in Fig. 1(a). The blue sheet and red sheet correspond to the eigenstate with gain and loss, respectively. An EP can be found at $g = 1$ and $\delta = 0$. We consider a loop enclosing the EP parameterized by $g(t) = 1 - \rho\cos(\gamma t)$ and $\delta(t) = \rho\sin(\gamma t)$, where the loop radius $\rho \leq 1$ and $\gamma$ measures the adiabaticity. When $t = \pi/\gamma$ (i.e., $g = 1+\rho$, $\delta = 0$), the two eigenvectors are solved to be

$$|\psi_A\rangle = \left[1, i\left(\rho+1+\sqrt{\rho^2+2\rho}\right)\right]^T, \quad |\psi_B\rangle = \left[1, i\left(\rho+1-\sqrt{\rho^2+2\rho}\right)\right]^T,$$

indicating that the eigenstates are in the *PT*-broken phase. However, the real parts of the eigenvalues are found to bifurcate (see Fig. 1(a)), which is completely opposite to that in *PT*-symmetric systems where the real parts coalesce in *PT*-broken phase [7]. This is the key difference between *PT*-symmetric systems and anti-*PT*-symmetric systems [36,37], and will result in different dynamics when the EP is dynamically encircled.

We consider a loop with the starting point and end point at $t_0 = -\pi/|\gamma|$ and $t_{end} = \pi/|\gamma|$, respectively. The loop is counter-clockwise when $\gamma < 0$ and clockwise when $\gamma > 0$. We first investigate counter-clockwise loops with $\rho = 0.5$ and $\gamma = -0.5$. We solve the time-dependent equation numerically and extract the amplitudes of the instantaneous eigenstates at each time step, i.e., $|\psi(t)\rangle = C_G(t)|\psi_G(t)\rangle + C_L(t)|\psi_L(t)\rangle$, where the subscript G and L are associated with the eigenstate on the gain sheet (Im($\lambda$) > 0) and loss sheet (Im($\lambda$) < 0), respectively. The obtained amplitude coefficients $C_G$ and $C_L$ for the process with state A and state B being the initial state are shown in Figs. 1(c) and 1(d), respectively, which can be used to draw the trajectory of the state evolution on the energy Riemann sheets, as shown by the white and yellow curves in Fig. 1(a). The trajectory is marked on the gain (loss) sheet colored in blue (red) when $|C_G| > |C_L|$ ($|C_G| < |C_L|$). The yellow trajectory shows a process that the state evolves entirely on the gain sheet (blue color). We see the initial state B gradually transforms to state A in this stable process (also see Fig. 1(d)). The situation is quite different when the initial state is state A (see the white trajectory in Fig. 1(a)). The state at first propagates on the loss sheet (red sheet) where the state is not stable. A NAT occurs after some time and the state jumps to the gain sheet on which it stays for the rest of the loop. As a result, the final state is still state A, the same as the initial state because of the NAT (also see Fig. 1(c)). These two processes indicate that the final state for counter-clockwise loops is always state A, regardless of the initial state. We apply the same analysis to study clockwise loops. The results with $\rho = 0.5$ and $\gamma = 0.5$ are plotted in Figs. 1(b), 1(e) and 1(f), showing that the final state for clockwise loops is always state B, in contrast to that of counter-clockwise loops. Such dynamical behavior is called chiral dynamics, i.e., encircling the EP in different directions results in different final states but independent of the initial state.

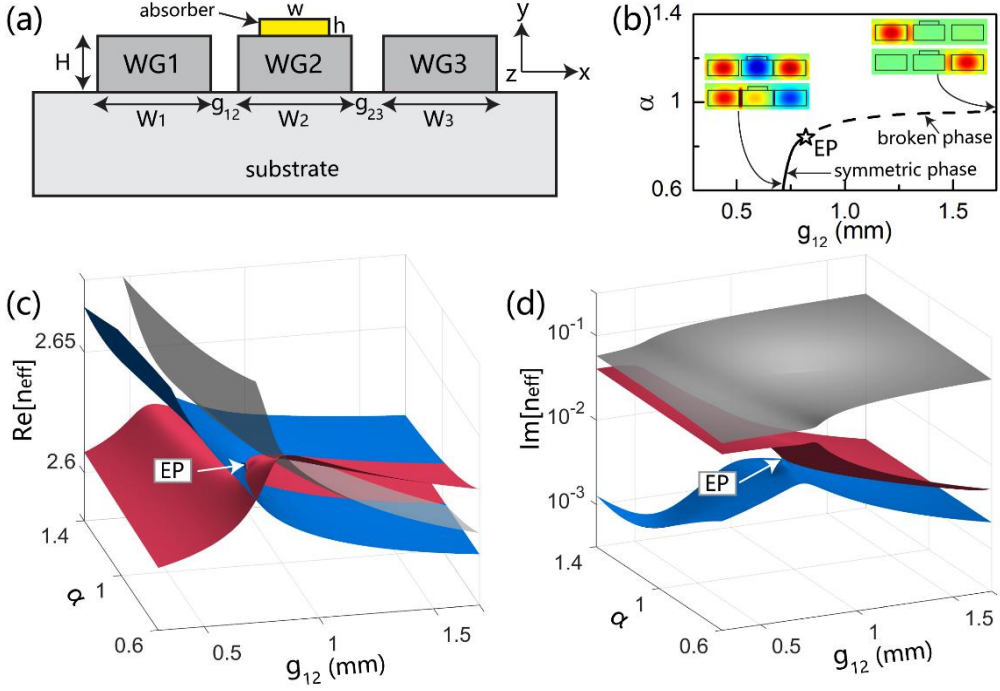

FIG. 2. (a) Cross-sectional views of the coupled waveguides with an absorber placed on top of waveguide-2. (b) Parameter space of the system. The star marks the EP and the inset shows the $E_x$ field distributions of the two eigenmodes on the blue and red energy sheet. (c) Real part and (d) imaginary part of the effective mode index of the system as a function of $g_{12}$ and $\alpha$.

The chiral dynamics can be understood mathematically using the method introduced in Ref. [26]. Inserting the expression of $g(t)$ and $\delta(t)$, the time-dependent Hamiltonian described by Eq. (1) can be written as a pair of coupled first-order differential equations for $a(t)$ and $b(t)$,

$$\begin{cases} \dfrac{\partial a(t)}{\partial t} + i\left(\rho e^{i\gamma t} - 1\right)a(t) + b(t) = 0 \\ \dfrac{\partial b(t)}{\partial t} - i\left(\rho e^{i\gamma t} - 1\right)b(t) + a(t) = 0 \end{cases}, \quad (2)$$

which can be rewritten as a second order differential equation for $a(t)$,

$$\dfrac{\partial^2 a(t)}{\partial t^2} + \left[\rho^2 e^{2i\gamma t} - \rho(2+\gamma)e^{i\gamma t}\right]a(t) = 0. \quad (3)$$

This equation can be transformed into a degenerate hypergeometric differential equation and the solutions are confluent hypergeometric functions. We first consider counter-clockwise loops with $\gamma < 0$. The solutions at time step $t$ can be related to the initial condition by

$$\begin{bmatrix} a(t) \\ b(t) \end{bmatrix} = \sigma(t) M_1(t) M_2 M_3 \begin{bmatrix} a(t_0) \\ b(t_0) \end{bmatrix}, \quad (4)$$

where $\sigma(t) = i\Gamma(-1/\gamma)e^{\rho(1-e^{i\gamma t})/\gamma}$ and the matrices are

$$M_1(t) = \begin{bmatrix} F^{(0)} & U^{(0)} \\ iF^{(0)} + 2\rho i e^{i\gamma t}F^{(1)}/\gamma & iU^{(0)} - 2\rho i e^{i\gamma t}U^{(1)}/\gamma \end{bmatrix}, \quad (5a)$$

$$M_2 = \begin{bmatrix} i\rho U^{(0)}_{t=-\pi/|\gamma|}/\gamma - 2i\rho U^{(1)}_{t=-\pi/|\gamma|}/\gamma^2 & -U^{(0)}_{t=-\pi/|\gamma|} \\ -i\rho F^{(0)}_{t=-\pi/|\gamma|}/\gamma - 2i\rho F^{(1)}_{t=-\pi/|\gamma|}/\gamma^2 & F^{(0)}_{t=-\pi/|\gamma|} \end{bmatrix}, \quad (5b)$$

$$M_3 = \begin{bmatrix} 1 & 0 \\ i(1+\rho)/\gamma & -1/\gamma \end{bmatrix}, \quad (5c)$$

where $F^{(n)}$ and $U^{(n)}$ are confluent hypergeometric functions $F(n-1/\gamma, n+1, 2\rho e^{i\gamma t}/\gamma)$ and $U(n-1/\gamma, n+1, 2\rho e^{i\gamma t}/\gamma)$, respectively [39]. We define $M(t) = M_1(t)M_2 M_3$ and study the matrix elements $m_{ij}$ at $t_{\text{end}} = \pi/|\gamma|$. Taking the adiabatic limit $\gamma \to 0$ and using asymptotic expansion of the confluent hypergeometric functions, we obtain $m_{21}/m_{11} = m_{22}/m_{12} = i(\rho + 1 + \sqrt{\rho^2 + 2\rho})$, which implies

$$\left.\dfrac{b(t_{\text{end}})}{a(t_{\text{end}})}\right|_{\text{CCW}} = \dfrac{m_{21}a(t_0) + m_{22}b(t_0)}{m_{11}a(t_0) + m_{12}b(t_0)} = i\left(\rho + 1 + \sqrt{\rho^2 + 2\rho}\right). \quad (6)$$

Equation (6) indicates that no matter what the initial state is, the final state is always state A for counter-clockwise loops.

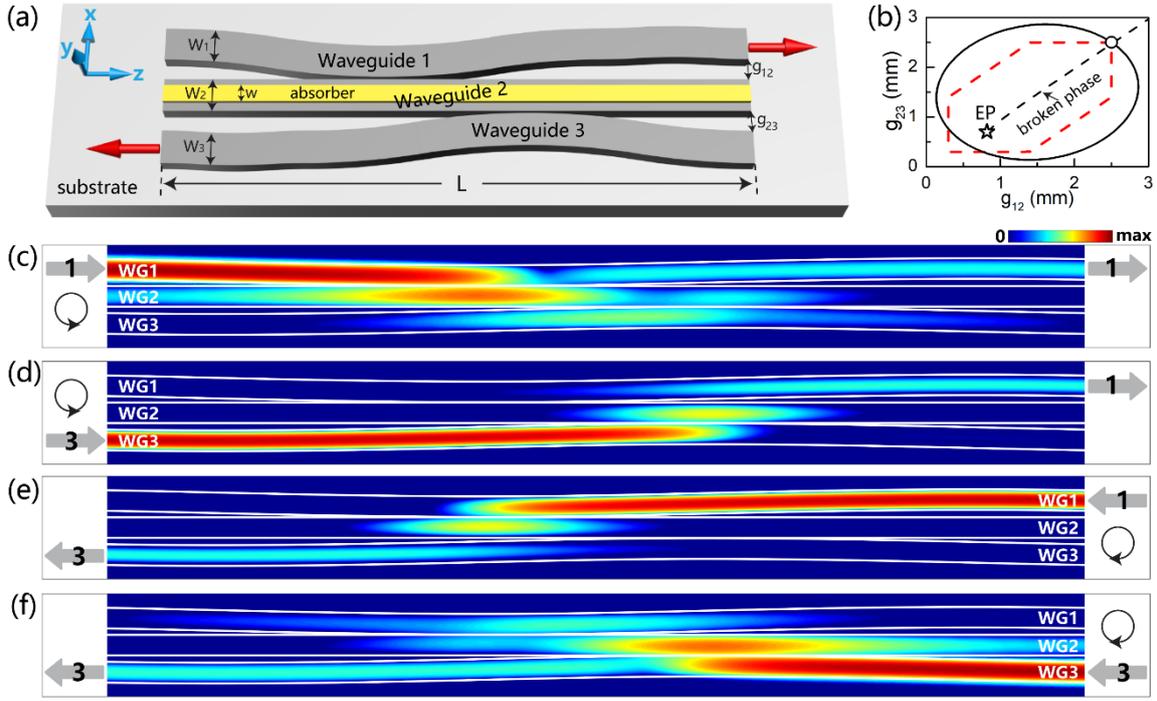

FIG. 3. (a) Top view of the coupled waveguides with two gap distances varying continuously along the waveguiding direction, corresponding to the dynamical encircling of an EP in anti-*PT*-symmetric systems. The two red arrows indicate the asymmetric mode switching, i.e., power always exists the system via waveguide-1 with left-hand side injections whereas via waveguide-3 with right-hand side injections. (b) Parameter space of the system. The circle and star mark the starting/end point and EP, respectively. The black solid curve and red dashed curve represent the trajectory of the EP encircling in numerical simulations and experiments, respectively, and their formulas are given in Supplemental Material. (c)-(f) Numerically simulated power flow distributions in the system with different encircling directions and injections (indicated in the inset).

Details of the above derivation can be found in Supplemental Material. For clockwise loops with $\gamma > 0$, we use a transformation $(a_{\text{CCW}}, b_{\text{CCW}}) \rightarrow (b^*_{\text{CW}}, a^*_{\text{CW}})$ which maps the counter-clockwise (CCW) loop solution to the clockwise (CW) loop solution (see Supplemental Material for details) so that

$$\left.\frac{b(t_{\text{end}})}{a(t_{\text{end}})}\right|_{\text{CW}} = \left(\left.\frac{a(t_{\text{end}})}{b(t_{\text{end}})}\right|_{\text{CCW}}\right)^* = i\left(\rho + 1 - \sqrt{\rho^2 + 2\rho}\right), \quad (7)$$

which indicates that the final state for clockwise loops is state B.

Equations (6) and (7) are the key results of our theory. They prove the chiral dynamics in anti-*PT*-symmetric systems when the starting point lies in the *PT*-broken phase. This is in sharp contrast to *PT*-symmetric systems, in which the dynamics is non-chiral with a starting point in the *PT*-broken phase [33], and only a starting point in the *PT*-symmetric phase can result in chiral dynamics [29]. The difference originates from the very different topological structure of energy Riemann surfaces of *PT*-symmetric and anti-*PT*-symmetric systems. For chiral dynamics to occur, the trajectory in the parameter space must start from a point where the two eigenstates carry the same imaginary part. This is characteristic of the *PT*-symmetric phase in *PT*-symmetric systems. In contrast, it is the *PT*-broken phase that has such properties in anti-*PT*-symmetric systems. Table I summarizes the different dynamics in the two systems. The chiral behavior was employed for asymmetric mode switching using the symmetric and anti-symmetric modes in *PT*-symmetric systems [29]. Based on the same principle, the symmetry-broken states can also be employed for asymmetric mode switching in anti-*PT*-symmetric systems. We now demonstrate the asymmetric mode switching in a realistic system consisting of three waveguides with the cross section illustrated in Fig. 2(a). The dimensions of waveguide-1 and waveguide-3 have a slight detuning (i.e., $W_1 \neq W_3$), and they are coupled via a waveguide-2 which has an absorber attached (see the yellow region). For completeness, we first demonstrate using a simple model Hamiltonian that coupling the two waveguides (i.e., waveguide-1 and waveguide-3) through a lossy waveguide-2 can effectively produce a purely imaginary coupling and hence anti-*PT* symmetry [37]. Consider a model Hamiltonian satisfying the equation

$$\begin{bmatrix} \beta_1 & \kappa' & 0 \\ \kappa' & \beta_2 + i\gamma' & \kappa' \\ 0 & \kappa' & \beta_3 \end{bmatrix} \begin{bmatrix} \phi_1 \\ \phi_2 \\ \phi_3 \end{bmatrix} = E \begin{bmatrix} \phi_1 \\ \phi_2 \\ \phi_3 \end{bmatrix}, \quad (8)$$

where $\beta_1$, $\beta_2 + i\gamma'$, and $\beta_3$ denote the original eigenvalues of the three waveguides, and $\kappa'$ is a real value representing the coupling. Eliminating $\phi_2$, we obtain

$$\begin{cases} \beta_1 \phi_1 + \dfrac{\kappa'^2 \phi_1 + \kappa'^2 \phi_3}{E - \beta_2 - i\gamma'} = E\phi_1 \\ \beta_3 \phi_3 + \dfrac{\kappa'^2 \phi_1 + \kappa'^2 \phi_3}{E - \beta_2 - i\gamma'} = E\phi_3 \end{cases}. \quad (9)$$

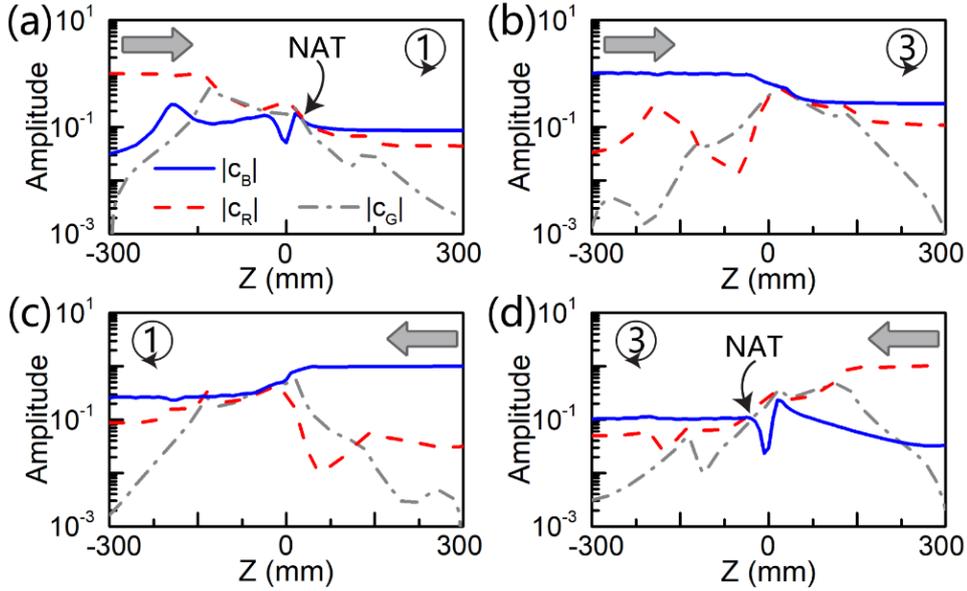

FIG. 4. (a)-(d) Extracted amplitudes of the instantaneous eigenmodes along the waveguiding direction with different encircling directions and injections (as indicated in the inset). The grey arrows mark the starting points of the loops.

For eigenvalues close to $\beta_2$, Eq. (9) can be simplified as

$$\begin{bmatrix} \beta_1 + i\dfrac{\kappa'^2}{\gamma'} & i\dfrac{\kappa'^2}{\gamma'} \\ i\dfrac{\kappa'^2}{\gamma'} & \beta_3 + i\dfrac{\kappa'^2}{\gamma'} \end{bmatrix} \begin{bmatrix} \phi_1 \\ \phi_3 \end{bmatrix} = E \begin{bmatrix} \phi_1 \\ \phi_3 \end{bmatrix}. \quad (10)$$

We note $\beta_1 \neq \beta_3$ since there is a detuning between waveguide-1 and waveguide-3. We define $\bar{\beta} = (\beta_1 + \beta_3)/2$ and $\Delta = (\beta_1 - \beta_3)/2$. Then the effective Hamiltonian describing the coupling between waveguide-1 and waveguide-3 becomes

$$H_{\text{eff}} = \begin{bmatrix} \Delta & i\dfrac{\kappa'^2}{\gamma'} \\ i\dfrac{\kappa'^2}{\gamma'} & -\Delta \end{bmatrix}, \quad (11)$$

after shifting the eigenvalues by a constant $\bar{\beta} + i\kappa'^2/\gamma'$. This effective Hamiltonian is anti-$PT$-symmetric.

To demonstrate the above analysis numerically, we used COMSOL [40] to calculate the effective mode index $n_{\text{eff}}$, i.e., the eigenvalue of the waveguide system as a function of two gap distances (i.e., $g_{12}$ and $g_{23}$) using structural parameters: $W_1 = W_2 = 8$ mm, $W_3 = 8.1$ mm, $H = 4$ mm, $w = 5$ mm and $h = 1$ mm. The permittivity of the waveguides and absorber are set to be 15.26 and $4+15i$, respectively. The background and substrate are assumed to be air and the frequency is 10 GHz. Although the parameter space introduced in the realistic system is different from that in the model Hamiltonian in Eq. (1), the dynamical behavior of encircling the EP is the same because the energy surfaces have the same topology. The real parts and imaginary parts of the eigenvalues are plotted in Figs. 2(c) and 2(d), respectively, where we define $\alpha = g_{23}/g_{12}$. The system supports three eigenmodes, which are represented by three Riemann sheets with different colors depending on the amount of loss. The grey eigenmode exhibits the highest loss and it hardly interacts with other two modes. We therefore focus on the red and blue Riemann sheets and find an EP located at $g_{12} = 0.82$ mm and $\alpha = 0.84$. Figure 2(b) shows the parameter space. The solid and dashed curve mark respectively the set of points in the 2D parameter space where the real and imaginary parts of the eigenvalues coalesce. The inset shows the $E_x$ field distributions of the eigenmodes residing on the blue and red sheet. On the solid curve, the two eigenmodes are found to be symmetric/anti-symmetric while they are symmetry-broken (i.e., one mode localized in waveguide-1 while the other one in waveguide-3) on the dashed curve. This indicates that the solid and dashed curve is in fact the $PT$-symmetric and $PT$-broken phase, respectively. It is then evident that the system is anti-$PT$-symmetric since the real/imaginary parts of the eigenvalues coalesce in the $PT$-symmetric/$PT$-broken phase.

Figure 3(b) redraws the parameter space as a function of the two gap distances. We consider a loop (the black solid loop) that encloses the EP with the starting/end point at $g_{12} = g_{23} = 2.5$ mm, which corresponds to a point in the broken phase with eigenmodes being symmetry-broken states. A schematic diagram of a waveguide system that can mimic the designed loop is illustrated in Fig. 3(a), where the two gap distances change continuously along the waveguiding direction (i.e., z-axis). The system length $L$ is set to be 600 mm and other structural parameters are kept the same as those in Fig. 2. Excitation of initial states on the left-hand side of the system leads to state evolutions following a counter-clockwise loop while incidence from the right-hand side leads to a clockwise loop. We calculated the wave transmission in the system and the desired initial state at the boundary was excited by a pre-calculation of the eigenmodes using the boundary mode analysis module of COMSOL [40]. The distributions of the z-component power flow for counter-clockwise loops are shown in Figs. 3(c) and 3(d), with the initial symmetry-broken state localized in

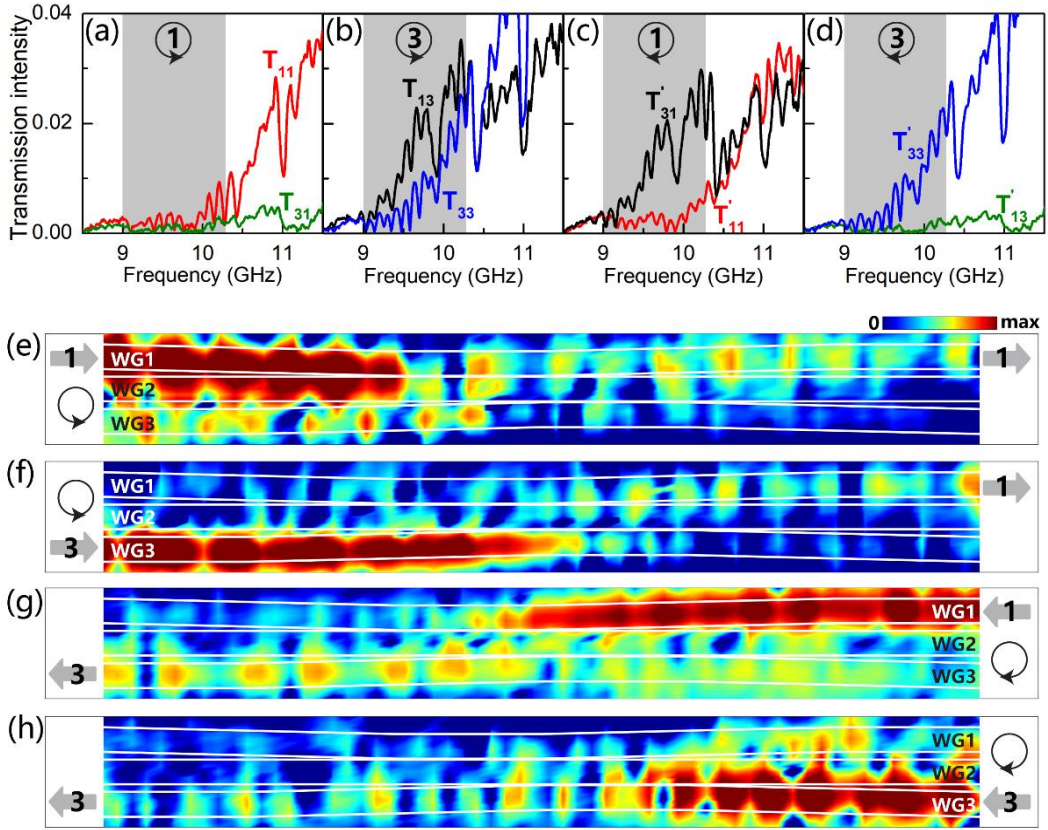

FIG. 5. (a)-(d) Experimentally measured transmission spectra with different encircling directions and initial states (indicated in the inset). The grey region marks the frequency range in which the phenomenon of asymmetric mode switching can be observed. (e)-(h) The corresponding measured electric field intensity distributions on top of the waveguide system at 9.6 GHz.

| System | Starting point in *PT*-symmetric phase | Starting point in *PT*-broken phase |
| --- | --- | --- |
| *PT*-symmetric systems | Chiral dynamics [29] | Non-chiral dynamics [33] |
| Anti-*PT*-symmetric systems | Non-chiral dynamics [this work] | Chiral dynamics [this work] |

Table 1 Comparison of the dynamical encircling of an EP in PT-symmetric systems and anti-PT-symmetric systems.

waveguide-1 and waveguide-3, respectively. We find that no matter via which waveguide the power is injected, the wave always exits the system via waveguide-1. The results for clockwise loops are shown in Figs. 3(e) and 3(f). In contrast, the final state is found to localize in waveguide-3 regardless of the initial state. This is a direct demonstration of the asymmetric mode switching for symmetric-broken states, which is different from that in *PT*-symmetric systems where the asymmetric mode switching applies to symmetric and anti-symmetric modes [29].

We investigate the dynamics by expanding the field profiles at each $z$ as a sum of the instantaneous eigenfields, i.e., $\mathbf{E}_t(z) = c_B(z)\mathbf{E}_B(z) + c_R(z)\mathbf{E}_R(z) + c_G(z)\mathbf{E}_G(z)$, where $\mathbf{E}_t$ is the transverse electric field, and the subscript B, R, and G denote the instantaneous eigenmode on the blue, red and grey Riemann sheet respectively. The amplitude coefficients $c_B$, $c_R$ and $c_G$ were determined by first constructing left eigenvectors and then doing a projection of the instantaneous fields onto the left eigenfields (see Supplemental Material for details). The calculated coefficients are plotted in Figs. 4(a)-4(d), corresponding to the dynamics in Figs. 3(c)-3(f), respectively. We find that for each encircling direction, the two processes with different initial states exhibit different dynamics. Specifically, one process is adiabatic as the blue curve dominates the entire trajectory (see Figs. 4(b) and 4(c)), which enables the wave to transform from one waveguide to the other one, corresponding to a state flip (see Figs. 3(d) and 3(e)). The other process has a NAT (at the crossing of the blue and red curves in Figs. 4(a) and 4(d)) and the final state is the same as the input (see Figs. 3(c) and 3(f)). These results well reproduce the dynamics obtained from the model Hamiltonian in Fig. 1. We also note from Fig. 4 that when the state approaches the end point, it always stays on the lower-loss blue sheet where it is more stable. The blue sheet is not continuous in the *PT*-broken phase of anti-*PT*-symmetric systems (see Figs. 1(a) or 2(c)) so that encircling the EP in different directions leads to different final states. This is the key reason of the chiral dynamics.

Microwave experiments were performed to demonstrate the asymmetric mode switching. The waveguides are made of yttrium iron garnet (YIG) with a relative permittivity of ~15.2. The trajectory of the experimental system is shown by the red dashed curve in Fig. 3(b). It is slightly different from that in numerical simulations since the system is composed of several straight YIG strips (see Supplemental Material for more experimental details). We measured the transmission spectra for different encircling directions and initial states using an Agilent Technologies 8720ES Network Analyzer. The results are shown in Figs. 5(a)-5(d), where $T_{ij}$ ($T'_{ij}$) represents the measured transmission intensity from waveguide-*j* to waveguide-*i* in a counter-clockwise (clockwise) loop. Since the system is designed for 10 GHz and the location of the EP varies with frequencies, the expected phenomenon should be observed in a specific range of frequencies. These frequencies are shaded in grey in Figs. 5(a)-5(d), where we find that the transmission corresponding to counter-clockwise loop is mainly dominated by $T_{11}$ (Fig. 5(a)) and $T_{13}$ (Fig. 5(b)), while that of clockwise loops by $T'_{31}$ (Fig. 5(c)) and $T'_{33}$ (Fig. 5(d)). This is an experimental observation of the asymmetric mode switching for symmetry-broken modes, i.e., the power always exits the system mainly via waveguide-1 in counter-clockwise loops and waveguide-3 in clockwise loops. The oscillations of the experimental spectra are due to Fabry-Pérot resonances as the system has a finite length of 600 mm. We also note $T_{ij} \approx T'_{ji}$ since the system is reciprocal. The electric field intensity distributions were measured on top of the waveguide system (~1 mm above the surface) with the help of a stepper motor. The results at ~9.6 GHz are shown in Figs. 5(e)-5(h), corresponding to the four cases in Figs. 3(c)-3(f). We find the experimental measurements can well reproduce the salient feature of numerical simulations, although some standing waves are present due to the Fabry-Pérot resonance. We note that the Fabry-Pérot effect is not observed in Figs. 3(c)-3(f) due to the matched boundary conditions used in numerical simulations.

In summary, we have shown both theoretically, numerically and experimentally that dynamically encircling an EP in anti-*PT*-symmetric systems can lead to a chiral behavior when the starting point lies in the *PT*-broken phase. The behavior depends on the starting point of the loop, and it is chiral when the two eigenstates at the starting point exhibit the same loss. As a result, the symmetry-broken states in anti-*PT*-symmetric systems can be used for asymmetric mode switching. Our work has revealed new physics of anti-*PT*-symmetric systems and may pave the way towards applications of symmetry-broken modes in non-Hermitian systems.

**ACKNOWLEDGEMENTS**

This work was supported by the Hong Kong Research Grants Council through grant no. AoE/P-02/12. X.-L.Z. was also supported by the National Natural Science Foundation of China (grant no. 61605056) and the China Postdoctoral Science Foundation (grant no. 2016M591480).